\documentclass[11pt]{article} 
\usepackage[utf8]{inputenc} 
\usepackage{geometry} 
\usepackage{graphicx} 

\usepackage{booktabs} 
\usepackage{array} 
\usepackage{paralist} 
\usepackage{verbatim} 
\usepackage{subfig} 
\usepackage{tikz}
\usepackage{changes}
\usepackage{bbm}
\usepackage{dsfont}
\usepackage{amsmath,amsthm,amsfonts,amssymb}
\usepackage{ulem}
\usepackage[active]{srcltx}
\usepackage{hyperref}
\hypersetup{colorlinks=true,pdfborder={0 0 0}}
\usepackage{tikz}
\usepackage[affil-it]{authblk}
\usepackage{fancyhdr} 
\usepackage{sectsty}
\allsectionsfont{\sffamily\mdseries\upshape} 

\usepackage[nottoc,notlof,notlot]{tocbibind} 
\usepackage[titles,subfigure]{tocloft} 

\def \be{\begin{equation}}
\def \ee{\end{equation}}
\def\id{\mathbbm{1}}
\def\half{\mbox{$\frac 1 2$}}

\newcommand{\blue}[1]{{\color{blue} #1}}
\newcommand{\bra}[1]{\left\langle #1 \right|}

\newcommand{\ket}[1]{\left| #1 \right\rangle}
\newcommand{\der}[2]{\frac{d #1}{d #2 }}
\newcommand{\e}{\varepsilon}

\newcommand{\mbf}[1]{\mathbf{ #1} }
\newcommand{\average}[1]{\mathbb{E}\left( #1\right)}
\newcommand{\C}[1]{{\cal{ #1 }}}

\def\lin{{\cal L}}
\DeclareMathOperator{\sgn}{sgn}
\DeclareMathAlphabet\mathbfcal{OMS}{cmsy}{b}{n}
\numberwithin{equation}{section}
\newtheorem{exa}{Example}[section]
\newtheorem{rem}{Remark}[section]
\newtheorem{claim}{Claim}[section]

\begin{document}


\author[1] {J.E. Avron}
\author[1]{O. Kenneth}
\affil[1]{Dept. of Physics, Technion, Israel} 
\author[2]{A. Retzker}
\affil[2]{Racah Institute of Physics, The Hebrew University, Israel}
\author[1]{M. Shalyt}


\title{Lindbladians for controlled stochastic Hamiltonians  }
\maketitle
\begin{abstract}
We construct  Lindbladians associated with {\em controlled}
{stochastic Hamiltonians} in weak coupling.  This allows to determine the power spectrum of the noise from measurements of dephasing rates; to optimize the control and to 
test numerical algorithms that solve controlled stochastic Schr\"odineger equations.
A  few examples are  worked out in detail.

\end{abstract}

	 \color{black}

\section{The problem and the result}

{This article  describes} Lindbladians  associated with {\em controlled} stochastic Hamiltonians in weak coupling. Controlled stochastic Hamiltonians arise  in the context of ``dynamical decoupling''  and ``coherent control''   and are used to examine protocols 
for extending the coherence {time }of qubits \cite{kurizki,lidar,alicki-lidar}. 

Lindbladians in the weak coupling limit  have been rigorously studied in 
\cite{davies,Lindblad, GoriniKossakowski,BreuerPet,davies-markovian,LebowitzSpohn,alicki,salgado,morigi}   
in the {\it  time independent} setting. 
Recently,   controls aimed at  extending the coherence of qubits  have been suggested in \cite{kurizki,lidar} and periodically  controlled Lindbladians have been {studied} in  \cite{alicki-lidar,Szczygielski}. However a careful derivation of the Lindbladians for the controlled stochastic evolutions  and in particular  Eq.~(\ref{main-o}) for {\em general} and in Eq.~(\ref{Fcg}) for {\em stationary control} are new. This is also the case for the notion of effective control introduced  in Appendix \ref{effective} and some of the examples in section \ref{s:ex}.     

Consider the  stochastic  controlled Hamiltonian\footnote{Controlled stochastic adiabatic evolutions are studied in \cite{fraas}.}.
\be\label{Hlab}
\underbrace{\left( \sum \xi_\alpha(t)  H_\alpha\right)}_{weak~noises}+\underbrace{H_c(t)}_{control}, 
\ee
$H_\alpha$ are fixed  Hermitian   matrices representing independent and in general non-commuting sources of noise.  
$\xi_\alpha$ are stationary Gaussian random processes
\be\label{JJ}
\average{\xi_\alpha(t)}=0,\quad\average{\xi_\alpha(t)\xi_\beta(u)}=J_{\alpha\beta}\left({|t-u|}\right)\,. 
\ee
with  $J$  rapidly  decreasing on {a} time scale $\tau$.
We  {shall sometimes} assume, w.l.o.g., that  $J$ is a diagonal matrix (this can be achieved by a   redefinition  of $H_\alpha$).  
{A spin in a magnetic field having fixed direction but noisy amplitude, often a good approximation \cite{kurizki,lidar}, 
 is represented by  a single term $\alpha$.}
The case where the direction of the field  is  also stochastic is modeled by several $\alpha$'s and gives rise to noise that is non-commutative (not been treated before.)
$H_c$, a  {\em  time-dependent} (Hermitian) matrix, represents the control.


It is convenient to reformulate the problem  in the interaction picture. Let 
 \be \label{model}
H^I_\xi (t)=\sum_\alpha \xi_\alpha (t)H^I_{\alpha}( t) ,\quad H^I_{\alpha}( t)=  
V^*( t) H_\alpha  V( t)
 \ee
where $V( t)$ is the unitary generated by the control $H_c(t)$, 
	\be\label{H}
	   H_c= i  \dot V( t)V^*( t),\quad V(0)=\id
	\ee
 Weak coupling parameter in the present context is defined by
\be\label{e}
  \e^2={\tau \|\tilde J\|}\,\|H_\alpha\|^2\ll 1
\quad \tilde J(\omega)=\int_{-\infty}^\infty e^{i\omega t} J(t) dt  \ge 0
\ee
{$\e$ is the phase acquired by the wave function during one correlation time (in the absence of control).}
There are several ways to think about weak coupling: If we 
think of  $\|\tilde J\|,\tau =O(1)$ then weak coupling means what its names suggests, namely, that the noise is weak in the sense that $\|H_\alpha\|=O(\e)$. 
An alternate approach, which is also insightful, is to take $\|\tilde J\|, \|H_\alpha\|=O(1)$ and  then weak coupling  means short correlation time  $\tau=O(\e^2)$. 

 The noise affects the (average) state on the {\em coarse grained}  time scale\footnote{In contrast to evolution of the (unaveraged) state where time scales 
$O(\tau/\e)$ lead to effects of $O(1)$.} 
\be\label{s}
 s=\e^2  t/\tau
\ee

 Control problems are characterized by the rate of  rotation of $H^I_\alpha(t)$. For example, when the control $H_c$ is time independent, (constant control), $\omega=\|H_c\|$ while for periodic Bang-Bang,  where $H_c(t)$ is a (periodic) sequence of delta pulses, $\omega_c$ is the frequency of the bangs. This gives rise to a second dimensionless parameter $\omega_c \tau$.
Our analysis of the weak coupling limit holds independently of $\omega_c\tau$.
Dynamical decoupling requires however $\omega_c\tau\gtrsim 1$ where
the time  scale  of the control, $\delta t=O(1/\omega_c)$,  is not resolved on the coarse grained time  scale $s$.

By  {\em stationary controls} we shall mean that 
$H^I_\alpha(t)$  has a finite number of Fourier coefficients. It is convenient  to factor $\e$ so that the Fourier coefficients are $\tilde H_\alpha(\omega)$ are
\be\label{stationary}
H^I_\alpha(t)=\e \sum_{\omega\in F}\tilde H_\alpha(\omega)\, e^{i\omega t}, \quad \tilde H_\alpha(\omega)=\tilde H^*_\alpha({-\omega})
\ee
$F$ a finite set.


{When $\e \ll 1$  we shall show that the evolution is governed by  (complete) positivity preserving Lindbladian}\footnote{This is related to the procedure of ``adiabatic elimination"\cite{James_effe}.}
\be\label{repar}
	\der{\rho}{s}=\lin_\e \rho 
	\ee
Moreover, we shall show that, in the case of stationary control, $\lin_\e$ has a limit as $\e\to 0$ given by:
\be
\lin=\sum _\alpha \lin_\alpha , \quad \lin_\alpha= {\mathbfcal H}_\alpha-{\mathbfcal D}_\alpha
\ee
 with 
\begin{align}\label{Fcg}
{\mathbfcal H}_\alpha \rho&=\frac {i\tau} 4 \sum_{\omega\in F} \tilde K_\alpha(\omega) 
\big [[\tilde H_\alpha(\omega),\tilde H_\alpha^*(\omega)],\rho\big],\nonumber \\  
{\mathbfcal D}_\alpha\rho&=\frac \tau 8\sum_{\omega\in F} \tilde J_\alpha (\omega)\big[ \tilde H_\alpha(\omega) , 
[\tilde H_\alpha^*(\omega),\rho]\big]
	\end{align} 
$\tilde J$ denotes Fourier transform and  $K$ is the anti-symmetric partner of $J$ :
\be
K_\alpha (u)=i \sgn(u)J_\alpha(u) ,
\ee
{Note that  $\tilde K(\omega)$ is real and $\tilde K(0)=0$.
${\mathbfcal H}_\alpha$ is a generator of unitary evolution since $[ \tilde H_\alpha(\omega) , \tilde{H}_\alpha^*(\omega)]$ is hermitian.  Since $\tilde J(\omega)\ge 0$ 
${\mathbfcal D}_\alpha\ge 0$ is a contraction, generating decoherence. }

\begin{rem}
The special form of $\lin_\alpha$ reflects the fact that stochastic evolutions are unital: The fully mixed state 
$\rho\propto {\id}$ is  stationary. 
\end{rem}
\begin{rem}
 Eq.~(\ref{Fcg}) can be used to determine $\tilde J$ from the measured
rates $\gamma_\alpha$ \cite{Ido,paola1,joerg1,bargil1}.  
See the examples in section \ref{s:ex}. Moreover, it implies that the optimal measurement time is $t=O(\tau/\e)$.   To see this observe that repeated measurements of a projection $P$ in  the state $\rho(t)$ generates a Poisson process with an average
	\[
	Tr(\rho(t)P)=p(\gamma,t)
	\]
Given total allotted time $T$, an optimal estimator minimizes the standard deviation  in $\gamma$. This fixes $t$ to be the minimizer of the sensitivity
\[ 
S= \frac{\sqrt{t p(1-p) }}{\vert \frac{dp}{d\gamma} \vert}
\]
Eq.~(\ref{Fcg}) determines $\rho(t)$ for $t\ge \tau/\e$.
For a depolarizing qubit 
\[
\rho(t)= e^{-\gamma t} P+(1-e^{-\gamma t})\half \id, \quad p(\gamma, t)=\frac{1-e^{-\gamma t}}2
\]
so $S$ takes its minimum at the left edge of the interval,  $t=O(\tau/\e)$.

\end{rem}

\color{black}
\section{Some exact results}
 
The Hamiltonian $H^I_\xi$ generates a stochastic  unitary evolution
$U_\xi$  given by\footnote{Since we are interested in the case $\tau >0$ we can avoid Ito's calculus.} 
  \begin{align}\label{U1}
	{U}_{\xi}(t) &= \left(e^{-i\int_0^t H^I_\xi(u) du}\right)_T\\
	&=\sum_{n=0}^\infty (-i)^n
	\int_{0\le t_1<t_2<\dots< t_n\le t}H_\xi(t_n) dt_{n}\dots H_\xi(t_1) dt_1 \nonumber
	\end{align}
The time ordering, denoted by the subscript $T$ in the first line  is defined explicitly in the second. 
More  crucial to us is the super-operator \footnote{We shall use script characters to denote super-operators.} $\C{U}_\xi$ acting on states
\be
\rho_0\mapsto \rho_\xi(t)= \C{U}_\xi \rho_0= U_\xi(t) \rho_0 U_\xi^*(t)
\ee
The super-operator can be written similarly
	  \begin{align}\label{Uxi}
	\C{U}_\xi (t)&= \left(e^{-i\int_0^t \C{H}_\xi(s) ds}\right)_T\\
	&=\sum_{n=0}^\infty (-i)^n
	\int_{0\le t_1<t_2<\dots< t_n\le t}\C{H}_\xi(t_n) dt_{n}\dots\C{H}_\xi(t_1) 
dt_1 \nonumber
	\end{align}	
where the super-operators $\C{H}$ acts by the adjoint action
 \be\label{ad}
	\C{H }\rho= \big(ad[H]\big)\rho\equiv [H,\rho]
	\ee
Note that $ ad[H_1] ad[H_2]\neq ad[H_1H_2]$. Rather, 
	\be\label{prod}
	(ad[H_1] ad[H_2])(\rho)= (\C{H}_1  \C{H}_2)(\rho)= \C{H }_1( \C{H }_2\rho)= 
[H_1, [H_2,\rho]]
	\ee
	We  also need the fact {that}
	\be\label{Jacobi}
 	ad\big[[A,B]\big]=\big[ad[A],ad[B]\big]
	\ee
which follows from Jacobi's identity.

The key object of  this study is  the (stochastic) averaged evolution 
 \be\label{average}
 \rho_0\mapsto \average{ \rho_\xi(t)}=\big(\C{U}(t)\big)\rho_0
 \ee
The super-operator $\C{U}(t)$  is trace preserving, (completely) positivity preserving and unital (i.e. $\C{U}\id=\id$), but, 
in general, not unitary or Markovian.

Recall that  for Gaussian averages 
\be\label{gaussian}
\average{e^{i\phi}}= e^{-\average{\phi^2}/2}  
\ee

It follows that for $\xi$ a stationary Gaussian process, 
\begin{align}\label{U2}
 \C{U}(t)&=\left(\exp\left(-\half \int_0^t du \int_0^t\, dv\,   \C{K}(u,v)\right)\right)_T\nonumber\\
&=\left(\exp\left(- \int_0^t du \int_0^u\, dv\,   \C{K}(u,v)\right)\right)_T
\end{align}
where
\begin{align}\label{K1}
\C{K}(u,v)=\average{\C{H}^I_\xi(u) \C{H}^I_\xi(v)}=
\sum_{\alpha\beta} J_{\alpha\beta}(u-v)\C{ H}^I_\alpha(u)\C{H}^I_\beta(v)
\end{align}

So far, no approximation has been made. However, the time ordering remains a  major complication\footnote{$\C{U}$ may 
be viewed as the grand canonical partition function of a 1-D quantum gas with short range interaction.}. 
For its precise meaning one can either go back to Eq.~(\ref{Uxi}), or alternatively, see the discussion  and 
graphical representation in Appendix \ref{s:lin}. There is no issue with time ordering  in two cases:  when $\xi$ is 
white noise and when the {(interaction picture)} Hamiltonian {commute} at different times.  We examine these cases first.

\subsection{White noise}\label{white}

White noise is the limit $\tau\to 0$  with $
\tau J=O(1)$.  By Eq.~(\ref{e}) this corresponds to  $ \e\propto \sqrt \tau\to 0$.   
Not surprisingly, the reduction of white noise to Lindblad evolution is exact.  
Since $J_{\alpha\beta}(t)=J_{\alpha\beta} \delta(t)$  we have
\begin{align}\label{Uwn}
 \C{U}(t)&=\left(\exp\left(\int_0^t du  \ \lin(u)\right)\right)_T
\end{align}
with
\begin{align}\label{Kwn}
\C{L}(t)=-\frac 1 2 \sum_{\alpha\beta}  J_{\alpha\beta}\C{ H}^I_\alpha(t)\C{H}^I_\beta(t)
\end{align}
{Since $\lin$ and $\C{ H}^I$ have the same time argument $t$,}  we may use the definition of time-ordering in Eq.~(\ref{Uxi}) with 
$\C{H}_\xi\mapsto \lin$, to conclude that $\lin$ is the generator of $\C{U}$.
The Lindbladian reduces to:
\be\label{lwn}
\C{L}_t\rho=-\frac 1 2 \sum_{\alpha\beta}  J_{\alpha\beta}\big[{ H}^I_\alpha(t),[{H}^I_\beta(t),\rho]\big]
\ee
The result is exact.    Since $H_\alpha(t)$ are unitarily related for different $t$ it follows that the 
family $\lin_t$ is unitarily related and  hence isospectral. In particular, the instantaneous dephasing rates are independent of 
the control $V(t)$. 
{This could be anticipated since to affect the dephasing rates, the control must be at least as fast as the noise correlations.}

\subsection{Commutative case}\label{s:commute}

In general, it is difficult to extract a generator of the evolution from Eq.~(\ref{U2}) because of the time ordering. 
In the commutative case this is not an issue and  the generator of the evolution follows from Eq.~(\ref{K1}). Let us denote
\begin{align}\label{K}
\C{G}(t)=-  \int_0^t\, du\,   \C{K}(t,u)_T&=-\sum_{\alpha}\int_0^t du  J_{\alpha\beta}(t-u)\left(\C{ H}^I_\alpha(t)\C{H}^I_\beta(u)\right)_T
\end{align}
We then have 
\begin{align}\label{GU}
 \C{U}(t)&=\left(\exp\left(\int_0^t du  \ \C{G}(u)\right)\right)_T
\end{align}
Of course, in the commutative case the index $T$ is redundant.  
Now although $\C{G}$ is {an exact generator}, it is not in general of {Lindblad} form. 
More precisely, it may fail to satisfy positivity at short times  as the following example shows.

\begin{figure}[htbp]
\begin{center}
\includegraphics[width=0.4\textwidth]{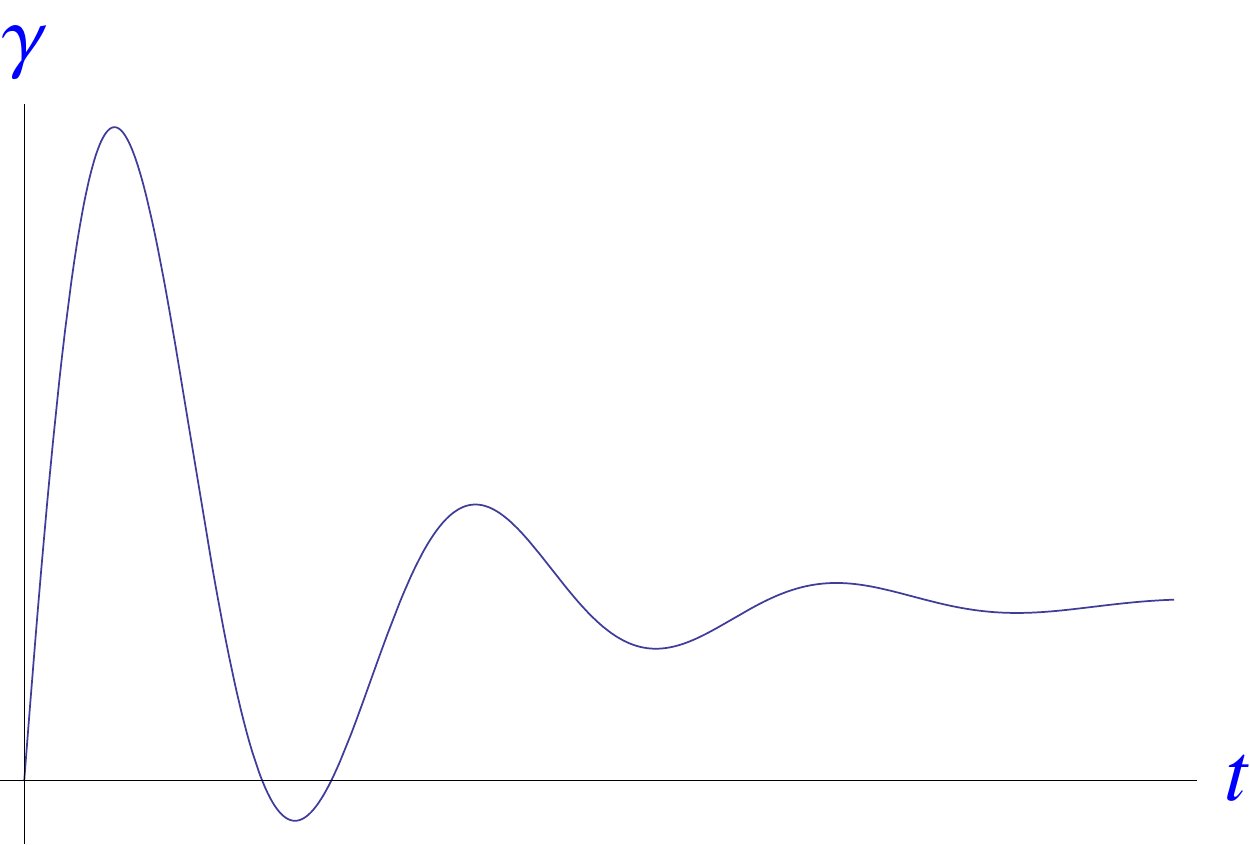}
\caption{The figure shows $\gamma(t)$ for $J(t)= e^{-|t|} \cos 4 t$ with $\gamma<0$ near the first minimum. \label{fig:gammaT} }

\end{center}
\end{figure}

\begin{exa}[Commutative case]\label{ex:com}
The commutative case arises, for example, when  the {(interaction picture)} noise has a stochastic amplitude but a fixed 
``direction'',  i.e. when
\[
H^I_\xi(t)= \xi(t)H_0
\]
 Since $H^I_\xi(t)$ is a commuting family, Eq.~(\ref{GU}) is exact and the generator of the evolution is
\be\label{stoc-lin}
\C{G}(t)\rho=-\frac {\gamma(t) } 2   \C{H}_0\C{H}_0\rho=-\frac {\gamma(t) } 2   [H_0,[H_0,\rho]],, 
\ee
The ``dephasing rate'' $\gamma(t)$  is given by 
\be\label{dotE}
\gamma(t)= 2\int_0^t du  J(t-u)=2 \int_0^t du  J(u)
\ee
 Although $\gamma(0)\ge 0$   for very short times (since $J(0)>0$), $\gamma(t)$ may be negative for   $t=O(\tau)$  
\footnote{Take e.g. $J(\omega)\propto \delta(\omega-\omega_0)+\delta(\omega+\omega_0) $}
as in Fig.~\ref{fig:gammaT}. 
In these cases  $\C{G}(t)$ does not generate a contraction  for all times.  This reflects the fact that the evolution 
is not strictly Markovian. At  longer times, $t\gg\tau$, one always has $\gamma(t)> 0$ (since $\tilde J(0)\ge 0$).

{Positivity} is regained in the weak coupling limit.  Here it is convenient to consider the limit in the sense of short correlation time so  
$\tau=\e^2$. We get from Eqs.~(\ref{s},\ref{repar})
\be
\lin=  \C{G}, \quad (\tau=\e^2)
\ee
 In the limit $\e\to 0$, we get for $s>0$,
\be\label{dir-noise}
 \lin \rho=-\frac { \tilde J(0)} {2}   [ H_0,[ H_0,\rho]], 
\ee
with time independent  positive dephasing rate:
\be
\quad 0<\tilde J(0)=\int_{-\infty}^{\infty} J(u) du=\lim_{\e\to 0}2\tau\int_{0}^{s/\e^2} J(u\tau) du 
\ee

\end{exa}    


\section{Weak coupling}\label{s:wc}

Our aim is to obtain an approximate generator {that is} valid  for small nonzero $\e$. 
The first step is to show that $\C{G}$ {defined in Eq.~(\ref{K})}  {remains} an approximate generator {in the noncommutative case }. More precisely, moving the time ordering $T$ in the exact 
formula Eq.~(\ref{U2}) into the exponential, comes with the penalty:
\begin{align}\label{U1}
 \C{U}(t)&=\left(\exp\left(-\half \int_0^t du \int_0^t\, dv\,   \C{K}(u,v)\right)\right)_T\nonumber\\
&=\exp\left(\int_0^t du \, \C{G}(u)\right)_{T_u}+O\left(\frac {\e^4 t}{\tau}\right)
\end{align}
{Here $T_u$ denotes time ordering with respect to the integration variable $u$. This differs from the usual time ordering defined
with respect to the argument of the hamiltonian. The error term results from the inequivalence of the two types of ordering. It}
is proportional to $t$ and  hence is cumulative. It reflects the non-commutativity of the Hamiltonian at different times 
(there is no error in the commmutative case as we have seen in section \ref{s:commute}). In particular, on the coarse grained time scale 
$s=O(1)\Leftrightarrow t=O(\tau/\e^2)$ the error is $O(\e^2)$.
We  justify this estimate in Appendix \ref{s:lin}.

As we have seen (again in   section \ref{s:commute})  $\C{G}$ may not be a Lindbladian. 
Our next step is to show that within the framework of weak coupling, $\C{G}$ is close to a 
generator that is  of the Lindbladian form up to an $O(\e^2)$ error. 

To show this we introduce a useful representation of the noise $\xi$  in terms
of white noises $W_\alpha$ :
\be\label{jj}
\xi_\alpha(t)= \int_{-\infty}^\infty du \, j_{\alpha\beta}(t-u) W_\beta(u)
\ee  
(summation implied) where
\be 
\average{W_\alpha (t)}=0,\quad 
\average{W_\alpha(t)W_\beta(u)}=\delta_{\alpha\beta}\delta(t-u)
\ee
There is freedom in defining $j$ which allows us to assume, w.l.o.g.,  that its Fourier transform is non-negative, $\tilde j(\omega)\ge 0$. 
$J$ is then the convolution of $j$ with itself:
	\be\label{convolution}
	 J_{\alpha\beta}(u-v)=\int_{-\infty}^\infty dw\,  j_{\alpha\gamma}(u-w)j_{\beta\gamma}(v-w)
	\ee
\begin{figure}[htbp]
\begin{center}
\includegraphics[width=0.4\textwidth]{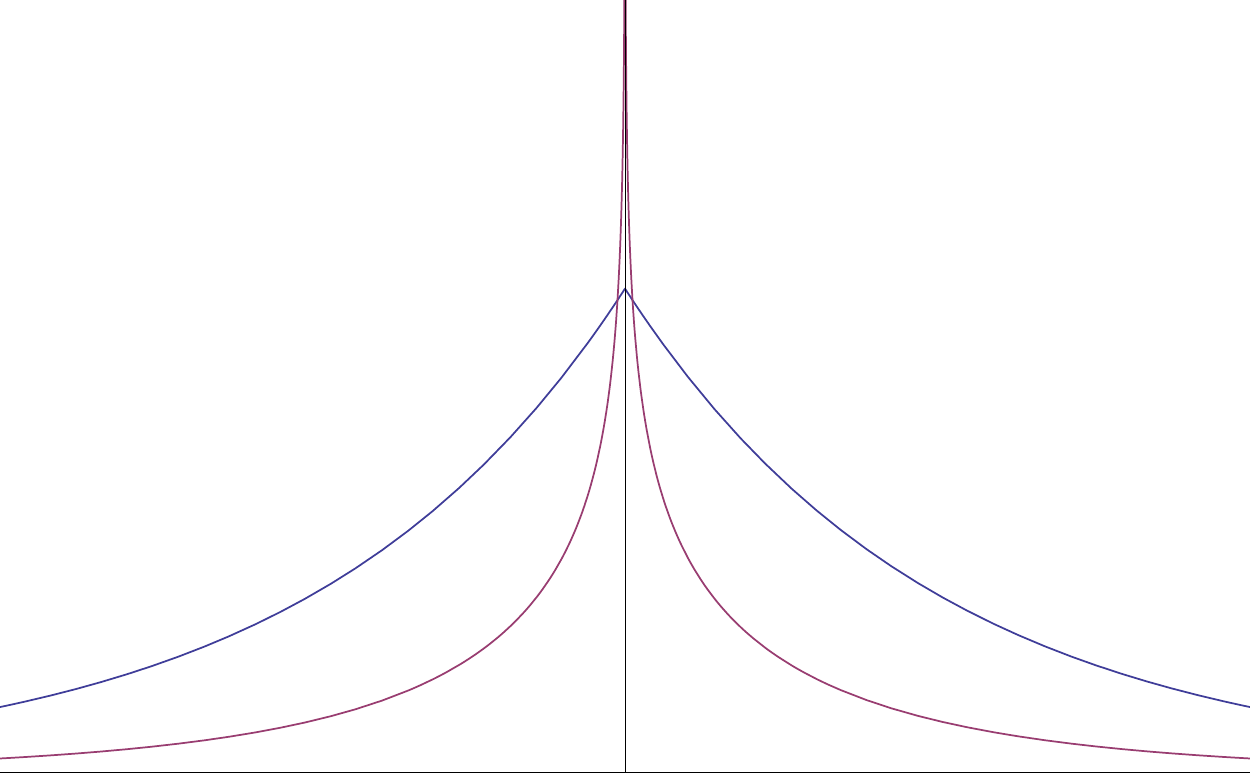}
\caption{$J(t)=e^{-|t|}$ and the corresponding  $j(t)=K_0(|t|)$, a Bessel 
function.  $j$ is narrower than $J$.}
\end{center}
\end{figure} 

With this notation in place we first note the identity
\begin{align}\label{error}
\left(\int_0^t dv  \int_0^t du\, \C{K}(u,v)\right)_T&=
\left(\int_0^t dv  \int_0^t du J_{\alpha\beta}(u-v)\C{ H}^I_\alpha(u)\C{H}^I_\beta(v)\right)_T\nonumber \\
&= \sum_\gamma\int_{-\infty}^{\infty}\, dw\,\left( 
\int_0^t\, du\,  j_{\alpha\gamma}(w-u)\C{H}_\alpha^I(u)\right)^2_T
\end{align}
which follows from Eq.~(\ref{convolution}). 

\begin{figure}[h]
\begin{center}
\begin{tikzpicture}
\draw [<->] (0,3) -- (0,0) -- (4,0);
\draw [blue] (-1/2,-1/2+1/4)--(3-1/4,3);
\draw [blue] (-1/2,-1/2-1/4)--(3,3-1/4);
\draw [red] (-1/2,2) --(3,2);
\draw [red] (2,0) --(2,3);
\node [below,blue] at (4,0) {v};
\node [left,blue] at (0,3) {w};
\node [above left,blue] at (0,2) {t};
\node [below,blue] at (2,0) {t};
\draw [blue, fill=blue] (2,2) -- (2+1/4,2) -- (2,2-1/4) -- (2,2);
\draw [blue, fill=blue] (0,0) -- (0-1/4,0) -- (0,0+1/4) -- (0,0);
\draw [blue, fill=green] (2,2) -- (2,2+1/4) -- (2-1/4,2) -- (2,2);
\draw [blue, fill=green] (0,0) -- (0+1/4,0) -- (0,0-1/4) -- (0,0);
\end{tikzpicture}
\caption{The figure illustrates the error terms in Eq.~(\ref{error2}) due to the change of domain of integration in Eq.~(\ref{error}). 
The error terms are represented by the blue and green triangles. The width of the strip bounded by the blue lines is $\tau$. 
Note that the figure is two dimensional while the actual domain of integration is three dimensional. \label{err} }
\end{center}
\end{figure}
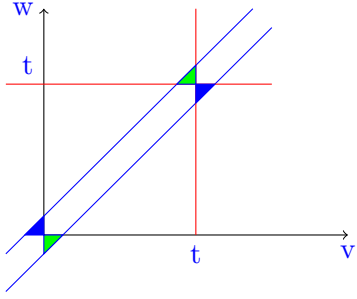

The second step is the claim that we  can interchange the limits of the $dw$ and $du$ integration in Eq.~(\ref{error}) 
up to a small error, i.e.
\begin{align}\label{error2}
\int_{-\infty}^{\infty}\, dw\,\left( \sum_{\alpha}\int_0^t\, du\,  j_{\alpha\gamma}(w-u)\C{H}^I_\alpha(u)\right)^2_T=
\int_{0}^{t}\, dw\,\left( \C{D}_\gamma(w)\right)^2_T+O(\e^2)
\end{align}
where 
\be \label{Lin}
\C{D}_\gamma(w)=\frac 1 2 \sum_{\alpha}\int_{-\infty}^{\infty} du\, j_{\alpha\gamma}(w-u)\C{H}^I_\alpha(u)\ee
This follows from the fact that $j(u)$ is localized near the origin on {a} time scale $O(\tau)$.   
It is clear that the main contribution to the integral comes from the region where both $w,u\in[0,t]$.
{The error corresponds to contributions where $w,u$ are in an $O(\tau)$ neighborhood  of the interval endpoints (see Fig.~(\ref{err})).
As this region has volume $O(\tau^3)$ and the integrand is $O(j^2H^2)$ we conclude that} 
the error is of magnitude $\tau^3 j^2 H^2\sim\tau^2 J H^2\sim\e^2$, whereas the first term is of order 
$t \tau^2 j^2 H^2\sim t \tau J H^2\sim  t \e^2/\tau\sim s $ which dominates the error.  
This proves Eq.~(\ref{error2}) with an error uniform in time.

It follows that for $\e$ small the (time-dependent) super-operator
	\be\label{main-o}
	\C{G}_\e=- \frac 1 2 \sum_\alpha\left(\C{D}_\alpha(t)^2\right)_T 
	\ee
generates a CP map which is $O(\e^2)$ close to $\C{U}$.  

 To rephrase ${\cal G}_\e$  in terms of operators, rather than super-operators, use
\be\label{order}
2\left(\C{H}(u) \C{H}(v)\right)_T= \{ \C{H}(u),\C{H}(v)\}+\sgn(u-v) [\C{H}(u),\C{H}(v)]
\ee 
and the dictionary in  Eq.~(\ref{prod},\ref{Jacobi}) gives a time dependent generator
\begin{align}\label{F}
	 \C{G}_\e \rho&= {i} [{H}^{ren}(t),\rho]-{1\over2} \sum_\alpha[ {D}_\alpha(t) , [{D}_\alpha(t),\rho]]\\
H^{ren}(t)&=\frac i 4 \sum_\alpha\int j_\alpha(u) j_\alpha(v) \sgn (u-v)~[H^I_\alpha(t+u),H^I_\alpha(t+v)] \, dudv\nonumber
	\end{align} 
 Since the operators $D_\alpha(t)$ and $H^{ren}(t)$ are  self-adjoint  $\C{G}_\e$ is a bona-fide {time dependent} generator of a CP map.  

\subsection{Coarse graining: The Lindbladian in the $\e\to 0$ limit}

So far we kept $\e$ small but finite and allowed arbitrary time dependence of $H^I(t)$. This gives the time dependent generator of the previous section.
To properly define the limit $\e\to 0$, one should also specify the limiting behavior of the dimensionless parameter $\omega_c\tau$.
If $\omega_c\tau\to 0$ then $\tau$ is the smallest time scale and $\xi$ becomes effectively equivalent to white noise discussed in section~\ref{white}. The interesting case and the one relevant to 
dynamic decoupling is when $\omega_c\tau\geq O(1)$ (where $\omega_ct=\omega_c\tau s/\e^2\to \infty$).
In this limit Eq.~(\ref{F}) reduces to Eq.~(\ref{Fcg}). To see this note:

\begin{itemize}
\item Weak coupling may be interpreted as  $J,\tau,\omega_c=O(1)$ while $\|H_\alpha\|=O(\e)$.  Eq.~(\ref{stationary}) then  implies that $\tilde H_\alpha(\omega)=O(1)$.
\item The ansatz Eq.~(\ref{stationary}) says that the integral in  Eq.~(\ref{Lin})  reduces to the sum: 
\be
{\cal D}_\alpha(t)= \frac \e 2 \sum_{\omega\in F} \tilde j (\omega) e^{i\omega t}\, ad\big(\tilde H_\alpha(\omega)\big)
\ee
\item
The limit $\e\to 0$ means that
\be\label{weak}
\lim_{\e\to 0} e^{i\omega\tau s/\e^2}=\begin{cases} 0 & \omega \neq 0\\
							1 & \omega =0
\end{cases}
\ee
in the sense of distributions. 
\item 
The limiting Lindbladian generates the evolution on the time scale $s=\e^2 t/\tau$, it is related to ${\cal G}_\e$ by 
$\lin=\tau\e^{-2} {\cal G}_\e$. 
\end{itemize}
It follows that for the second term in Eq.~(\ref{F}) we get
\[
\frac \tau {2\e^2}  {\cal D}_\alpha^2(t)\underset{\e\to 0}{\longrightarrow} \frac \tau 8 \sum_{\omega\in F} 
\tilde j_\alpha(\omega) \tilde j_\alpha(-\omega) ad (\tilde H_\alpha(\omega)) ad (\tilde{H}_\alpha(-\omega))
\]
which is $ {\mathbfcal D}_\alpha$ of Eq. ~(\ref{Fcg}).
Similarly, for the first term  in Eq.~(\ref{F}) we get
\begin{align}
i\frac \tau{\e^{2}}\, ad(H^{ren})&\underset{\e\to 0}{\longrightarrow}\frac {i\tau} 4\sum_{\alpha\omega} \int 
j_\alpha(u) j_\alpha(v) \sgn (v-u)~e^{i\omega(u-v)}\,ad\big([\tilde H_\alpha(\omega),\tilde H_\alpha^*(\omega))]\big) \, dudv\nonumber
\end{align} 
The $u,v$ integration an be carried out explicitly to give
\be
 \int j_\alpha(u) j_\alpha(v) \sgn 
(v-u)~e^{i\omega(u-v)} \, dudv\nonumber=\int J_\alpha(u) \sgn(u) e^{i\omega u}= \tilde K_\alpha(\omega)
\ee 

\begin{figure}[htbp]
\begin{center}\label{oded}
\includegraphics[width=0.45\textwidth]{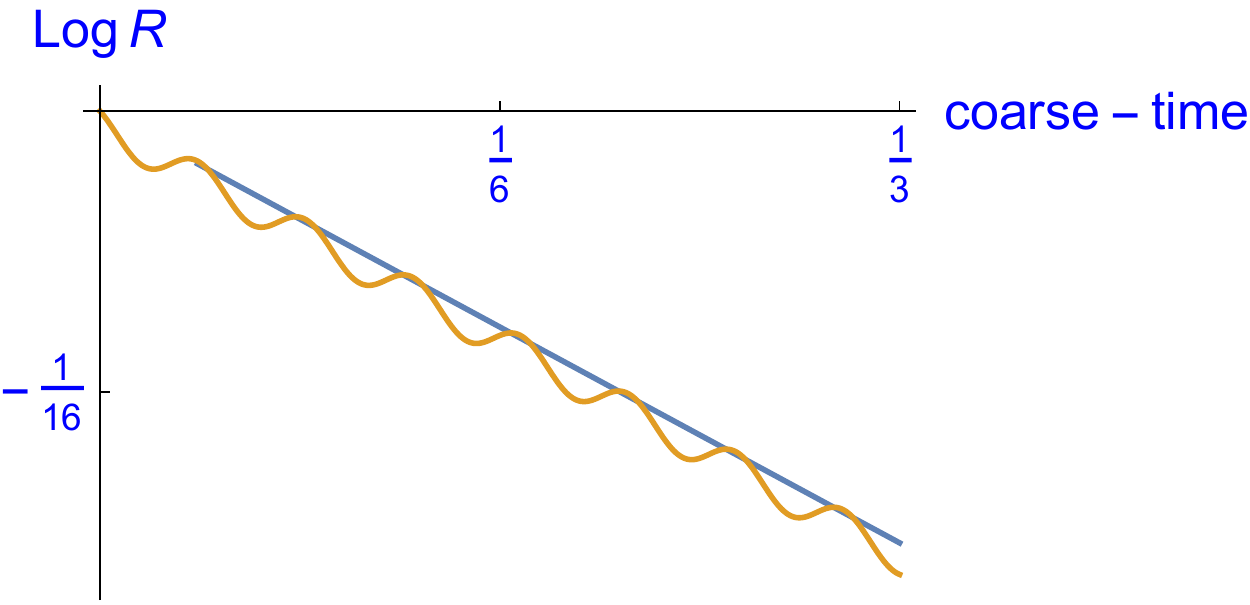}
\caption{For a qubit state $\rho= (\id+ \bold r\cdot \sigma)/2$ the figure shows $2\log |\bold r|$ as a function of coarse grained time $s$ for constant control. The parameters are the same as in Fig. \ref{gammaT}, i.e. $\e=0.15$ and $\omega_c\tau=\pi/2$.  The figure compares 
the average  Lindblad evolution on the course grained time scale, Eq.~(\ref{Fcg}),  and the time dependent Lindblad evolution, Eq.~(\ref{F}). Note the different scales in the two figures. The  oscillations are tiny.  }
\end{center}
\end{figure}  
\section{Examples}\label{s:ex}

\subsection{No conrol}\label{s:nc}
 Consider  the (commutative) stochastic Hamiltonian for spin $\mbf{S}$
		\be\label{spin}
		H= \xi  S_z
		\ee
Eq.~(\ref{dir-noise})  gives  the  dephasing Lindbladian 
	\be\label{dephasing}
	\lin\rho =- \frac{\gamma}2  [S_z,[S_z,\rho]],\quad \gamma=\tilde J(0)
	\ee
with a single rate parameter $\gamma$.  The coherences can be computed using the spectral 
properties of the super-operator of angular momentum given in Appendix \ref{a:spin}
\be\label{spectrum} 
Spectrum\bigg(\big(ad (S_z)\big)^2\bigg)=\big\{ m^2\, |\, m\in\{0,\dots, 2S\}\big\}
\ee
 The  coherence decreases quadratically with the polarization $m$:   
$[S_z,[S_z,\ket{m_1}\bra{m_2}]] =\ket{m_1}\bra{m_2}(m_1-m_2)^2$.

\subsection{Bang-Bang}\label{s:bb}
``Bang-bang" makes $\gamma$ smaller and improves the coherence: A sequence 
of rapid $\pi$ rotations about an axis perpendicular to the magnetic field self-average the noise. The $\pi$ rotations 
are given by the unitary:

\[V(t)=\begin{cases} e^{i\pi S_x} & \omega t\mod 2\pi \in (-\pi,0)\\
					{\id} & \omega t\mod 2\pi \in (0,\pi)\\	
	\end{cases}
\]
\color{black}
The controlled stochastic Hamiltonian corresponding to Eq.~(\ref{spin}) then takes the form:
		\be\label{Hbb(t)}
		H^I_\xi(t)=\xi(t)  S_z w(\omega t)
		\ee
where $w$ is the square wave
	\[w(t)=\begin{cases} 1 &  \omega t\mod 2\pi \in (-\pi,0)\\
					-1 & \omega  t\mod 2\pi \in (0,\pi)\\	
	\end{cases}
\]
Using the Fourier expansion 
\[W(t)=-\frac {2i} \pi \sum_n \frac{ e^{i (2n+1) t}}{2n+1}
\]
and Eq.~(\ref{Fcg}), we  obtain the functional form of the dephasing  Lindbladian of Eq.~(\ref{dephasing})  but 
with a renormalized\footnote{For a monotonically decreasing  $\tilde{J}(\omega)$ one has 
$\gamma_{b}(\omega)<\gamma_{b}(0)=\gamma$.}  $\gamma\mapsto \gamma_b(\omega)$: 
\be\label{bb}
	\gamma_{b}(\omega)= 
\frac {8} {\pi^2} \sum_{n\ge 0} \frac{\tilde J((2n+1)\omega)}{(2n +1)^2}
	\ee
Since,  $\tilde J(\omega)\to 0$    as $\omega\to \infty$, in the limit $\gamma_{b}(\omega)\to 0$, 
and there is no loss of coherence.

\subsection{Constant control}\label{s:cc}
Consider the stochastic Hamiltonian with time-independent control
\be\label{zx}
H= \underbrace{\omega_c S_z}_{control}+\xi S_x
\ee
 The control is effective
 in the sense of Appendix \ref{effective}.  In the interaction picture the stochastic Hamiltonian has the form 
		\be\label{H(t)}
		H^I_\xi(t)=\xi(t)\left( S_x\cos \omega_c t  +S_y  \sin\omega_c t\right)
		\ee
The frequency set $F$ in Eq.~(\ref{stationary}) has two elements, $F=\{\pm \omega_c\}$ and
\[
\tilde H(\pm\omega_c)=\frac 1 2 (S_x\pm iS_y)
\]
we find, from Eqs.~(\ref{Fcg})  the Lindbladian 
\be\label{acc}
	\lin\rho =-\frac{i\tilde K(\omega_c)} {8} [S_z,\rho]- \frac{\tilde J(\omega_c)}2 \sum_{j\in x,y} [S_j,[S_j,\rho]]
	\ee
The two terms in $\lin$ commute. This follows from
the fact that $J_i\equiv ad(S_i),\;\;i=x,y,z$ give an $SU(2)$ representation. As in any such representaion 
 $J_x^2+J_y^2$ is invariant under rotation around the $z$-axis, one has $[J_z,J_x^2+J_y^2]$=0. This may also be verified
directly by calculating the commutators.

It follows that the first term in $\lin$ determines the imaginary part of the eigenvalues  while the second term 
determines the real part. The coherence is then determined by the spectrum of
\[
spectrum\big(\sum_{j=x,y} ad(S_j)ad (S_j)\big)= \begin{cases} \{0, 1^{(2)},2\}& S=1/2\\
\{0,1^{(2)},2^{(3)},5^{(2)},6\}& S=1\end{cases}
\]
and the index denotes multiplicities. 
{In the case of general $S$ the spectrum is $\{j(j+1) - m^2 \vert\;\; |m|\le j\le 2S\}$ as computed in Appendix \ref{a:spin}.}
In particular using Schur's lemma implies that $0$ is always a simple eigenvalue. It follows that the Lindbladian is depolarizing: 
The unique equilibrium state is the fully mixed state.

\color{black}
  
\begin{figure}[htbp]
\begin{center}
\includegraphics[width=0.4\textwidth]{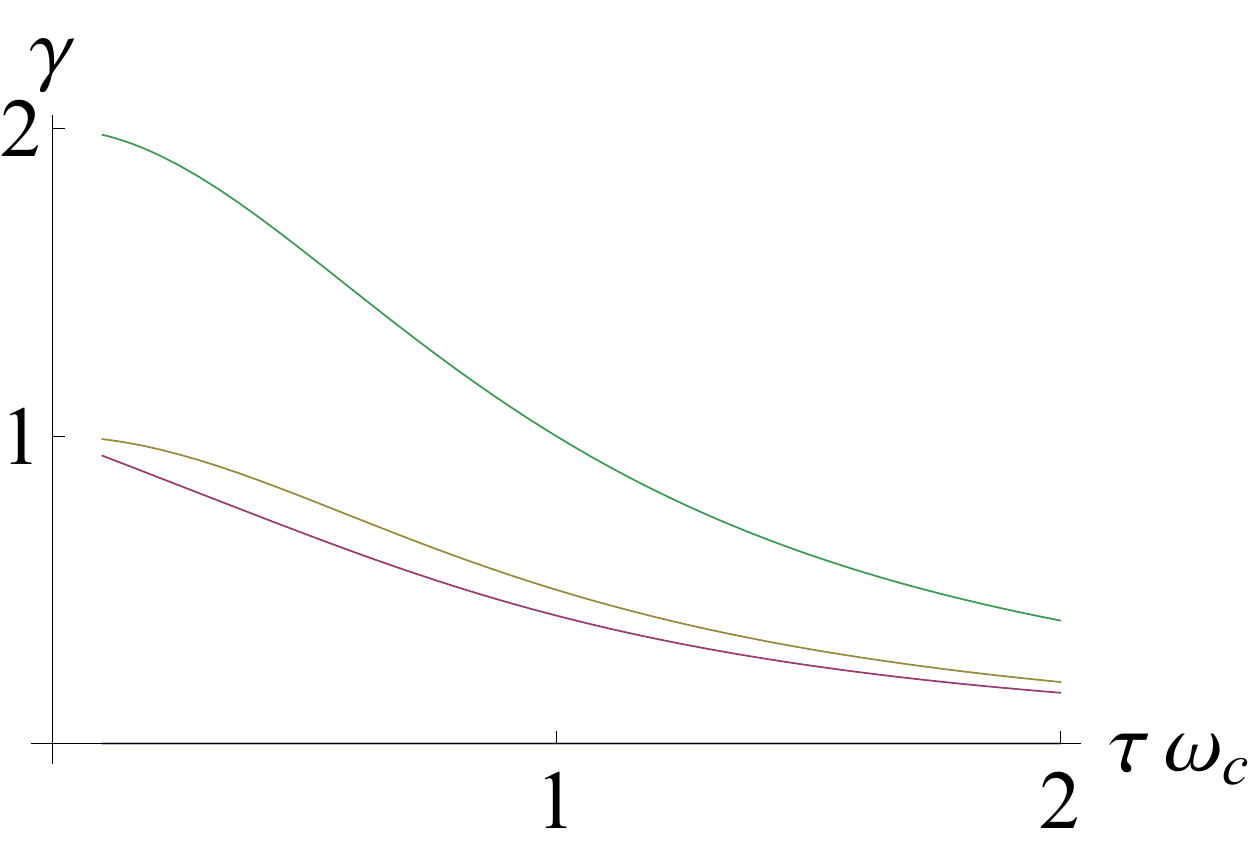}
\caption{The dephasing rate $\gamma$  for  $\tilde J(\omega_c)=(1+(\omega_c\tau)^2)^{-1}$ for a spin qubit in a 
stochastic magnetic field with a fixed direction as function of the rate of control $\tau\omega_c$.  The lowest  
line corresponds to Bang-bang of Eq.~(\ref{bb}). The two upper lines show the two nonzero eigenvalue of constant control. 
(Note that setting $\omega_c= 0$ in the graph is not meaningful  since  Eq.~(\ref{weak}) fails.)}
\end{center}
\end{figure}


\subsection{Non-commutative  noise}
The simplest case of non-commutative noise is ``planar'' noise 
	\[
	\sum_{\alpha=x,y} \xi_{\alpha} S_{\alpha} 
	\]
  Eq.~(\ref{Fcg})   gives the depolarizing Lindbladian
\be\label{gyg7}
	\lin\rho =- \frac{1}2 \sum_{j\in x,y}\gamma_j [S_j[S_j,\rho]] , \quad \gamma_j=\tilde J_j(0)
\ee
$H_c=\omega S_z$ is an effective control.  
Moreover, the results of  sections \ref{s:bb}, \ref{s:cc} carry over to this case, mutatis mutandis:{ Bang-Bang leads to Eq.~(\ref{gyg7}) with $\gamma\mapsto\gamma_b$ as in Eq.~(\ref{bb}).
Constant control leads to an equation similar to Eq.~(\ref{acc}) up to an extra factor of $2$.}
 

\subsection{Isotropic noise}

Isotropic noise is represented by the Hamiltonian 
	\[
	\sum_{\alpha=1}^3 \xi_{\alpha}  S_{\alpha}, \quad J_1=J_2=J_3 
	\]
leading to the isotropic depolarizing Lindbaldian
	\[
	\lin\rho =- \frac{\tilde J(0)}2 \sum_{j=1}^3 [S_j[S_j,\rho]], 
	\]
For $S=1/2$ constant control  is not effective
\footnote{For $S=1$ a possible effective constant control is $H_c=\sum\alpha_i S_i^2$.}.
One can, however, find an effective Bang-Bang.  
\footnote{The generalization to arbitrary spin is quite simple and only requires replacing  the $\sigma_k$ matrices in 
Eqs.~(\ref{uioger},\ref{oickm})
by the appropriate rotation operator $R_k=\exp(i\pi S_k)$.}

The simplest version of bang-bang about all three axes is associated with the unitary $V$
	\be\label{uioger}
	V(t)= \begin{cases}
	\sigma_1 & \omega t\mod 2\pi \in [0, \pi /2]\\
	\sigma_2 & \omega t\mod 2\pi \in [ \pi /2, \pi]\\
	\sigma_3 &  \omega t\mod 2\pi \in [ \pi,3\pi/2]\\
   {\id} &  \omega t\mod 2\pi \in [ 3\pi/2,2\pi]
	\end{cases}
	\ee
This  control self-averages the Hamiltonian in the interaction picture to zero but leads to a non-isotropic 
Lindblad equation (with $\gamma_1=\gamma_3\neq\gamma_2$).
In order to retain isotropy, we choose a somewhat more complicated $V(t)$ corresponding to
dividing $[0,2\pi]$ into 12 equal parts\footnote{We thank Ori Hirschberg for this suggestion.}. 
We demand $V(t)=V_j$ for $\omega t\mod 2\pi\in[2\pi j/12, 2\pi (j+1)/12]$ where
\be \{V_j\}_{j=1}^{12}=\{\sigma_1,\sigma_2,\sigma_3,{\id},\sigma_2,\sigma_3,\sigma_1,{\id},\sigma_3,\sigma_1,\sigma_2,{\id}\}
\label{oickm}\ee
This gives the stochastic Hamiltonian
\be
H_\xi=\sum\xi_\alpha S_{\alpha} w_\alpha(\omega t)
\ee
where $w_1(t)=w_2(t+2\pi/3)=w_3(t-2\pi/3)=\pm 1$ takes on $[2\pi j/12, 2\pi (j+1)/12]\mod2\pi$ the values

\be
w_1(t)\Leftrightarrow\{+1,-1,-1,+1,-1,-1,+1,+1,-1,+1,-1,+1\}_j
\ee

\begin{figure}[htbp]
\begin{center}
\includegraphics[width=0.4\textwidth]{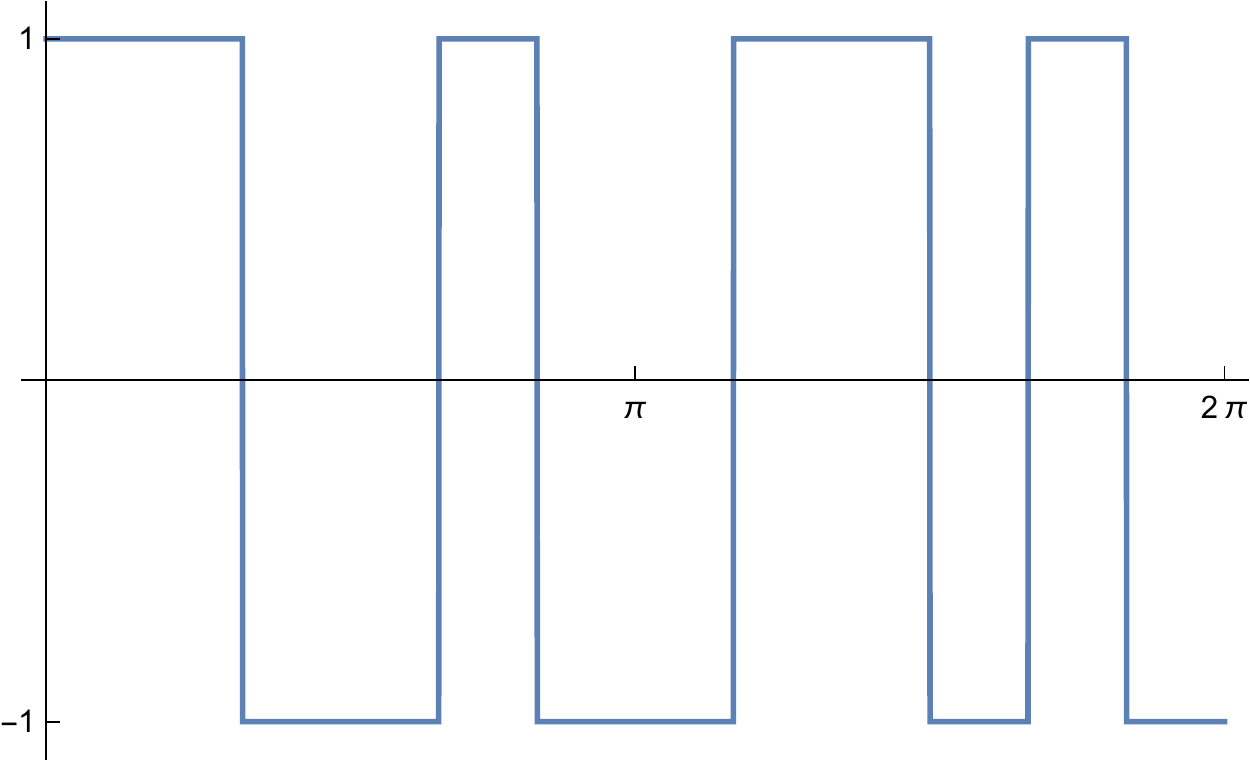}
\caption{The Bang-Bang control corresponding to the square wave  $w_1(\omega t)$ as function of time. 
It self-averages the noise Hamiltonian to zero while preserving the isotropy of the Lindbladian.  }
\end{center}
\end{figure} 
By symmetry considerations, ${\mathbfcal H}_\alpha=0$ (since $[\tilde H_\alpha(\omega),H_\alpha^*(\omega)]\propto [S_\alpha,S_\alpha]=0$).
and the depolarizing Lindbldian has renormalized rates: 
\[
	\lin\rho =- \frac{ \gamma(\omega)}2 \sum_{j=1}^3 [\sigma_j[\sigma_j,\rho]], 
	\] 
$\gamma(\omega)$ is a more complicated version of Eq.~(\ref{bb}) 
\be\label{iso}
	\gamma(\omega)=\frac 8 {\pi^2}\sum_{n\neq 0}\frac{\tilde J(\omega n)}{n^2}\sin^4\left(\frac{n\pi}{12}\right) p(n)
\ee
\[
p(n)= 
5+4\cos\left(\frac{n\pi}6\right)+2\cos\left(\frac{4n\pi}3\right)+(-1)^n\left(1+
4\cos\left(\frac{n\pi}2\right)+2\cos\left(\frac{2n\pi}3\right)\right)
\]

\color{black}

\subsection{Stochastic Harmonic oscillator}

The stochastic harmonic oscillator  provides a good model for trapped atoms, mechanical oscillators and trapped ions \cite{Milburn1}.
Since $\id$ is not a state in an infinite dimensional Hilbert space, the Lindbladian associated with stochastic evolution may have no 
stationary state. 

There are various types of noises one may consider.
The first is
\[   H_\xi=    \underbrace{\frac 12  \omega_c (p^2+x^2)} _{H_0}+\xi_p p+\xi_x x      \] 
with $\xi_x$ and $\xi_p$ Gaussian (possibly correlated) processes. This is known as `linear noise'  since it does not affect 
the frequency of the oscillator.

The interaction Hamiltonian is
\[
H_\xi^I= \xi_x\left( x \cos  \omega_c t + p\sin \omega_c t\right)+ \xi_p\left( p \cos  \omega_c t -x\sin \omega_c t\right)
\] 
$H_0$ is an effective control since $H^I_\xi$ has vanishing time average.
Observe that $ad(x)$ and $ad(p)$ commute  since
	\be\label{xp}
	[ad(x),ad(p)]=ad([x,p])=i\,ad(\id)=0
	\ee
It follows from Eq.~(\ref{Fcg})  that the Lindbladian is real (has no Hamiltonian piece) and has the form
\[
-2\lin =\Gamma_x\, ad(x) ad(x)+\Gamma_p\,ad(p) ad(p)  +2\Gamma_{xp} \{ad(x),ad(p)\}
\]
with matrix $\Gamma\propto \tilde J(\omega_c)$ at the oscillator frequency. Since $ad(x)$ and $ad(p)$ commute, 
and $spect\big(ad(x)\big)=spect\big(ad(p)\big)=(-\infty,\infty)$ and $\Gamma$ is a positive matrix, we have
\[
spect(\lin)=\{ -(\eta,\Gamma \eta)|\;\eta\in \mathbb{R}^2\}=(-\infty,0]
\]
 $0$ is in the spectrum but is not associated with an eigenvalue: There is no stationary equilibrium state.
 
 In the case of noise in the frequency of the harmonic oscillator 
 the Hamiltonian is:
 \[
H_\xi= \frac 12 \left((\omega_c+2 \xi_p) p^2+(\omega_c+2 \xi_x) x^2\right)=
\underbrace{\frac 12   \omega_c(p^2+x^2)} _{H_0}+\xi_p p^2+\xi_x x^2 
\] 
In the interaction picture one has
 \[
H^I_\xi= 
\xi_p( p\cos  \omega_c t -x\sin  \omega_c t)^2+\xi_x (x\cos  \omega_c t +p\sin  \omega_c t)^2 
\] 
Hence
\[
\tilde H_\alpha (0)=[\tilde H_\alpha(2 \omega_c),\tilde H_\alpha(-2 \omega_c)]=\frac 1 2 (p^2+x^2), 
\quad \tilde H_\alpha(\pm 2 \omega_c)=\frac 1 4(-1)^{\alpha} (p\pm i x)^2
\]
and
the Lindbladian is :
\be
\lin =  \underbrace{K_2 ~ad~ \tilde{H}(0)}_{unitary}+\underbrace{\Gamma_0\big(ad ~\tilde H(0)\big)^2}_{dephasing}+
\underbrace{\Gamma_2 \left(\big(ad~\tilde{H}(2\omega_c)\big)^2+\big(ad~\tilde{H}(-2\omega_c)\big)^2\right)}_{parametric~drive}
 \ee
 where $K_2\propto \tilde K( \omega_c)$, $\Gamma_0\propto \tilde J(0)$ 
and $\Gamma_2\propto \tilde J( 2\omega_c)$. 
{While all the terms $ \ket n \bra m$ are eigenstates of the dephasing part with eigenvalues $(m-n)^2$ the parametric drive part 
does not have a steady state and drives the system towards the infinite temperatures.}


\section{Comparison with stochastic evolutions}

Numerical algorithm for solving stochastic evolution equations have two advantages: They can work also beyond weak coupling and evolve states rather than density matrices. They also have several  disadvantage:  They tend to be slow because of the necessity to accumulating enough statistics; They are prone to long time drifts, and can be adversely affected by a poor random number generator and finally are prone to bugs.  Our results on the Lindbland evolutioon can be used to test numerical algorithms for stochastic evolutions in those cases that both apply. 

\begin{figure}[htbp]
\begin{center}
\includegraphics[width=0.9\textwidth]{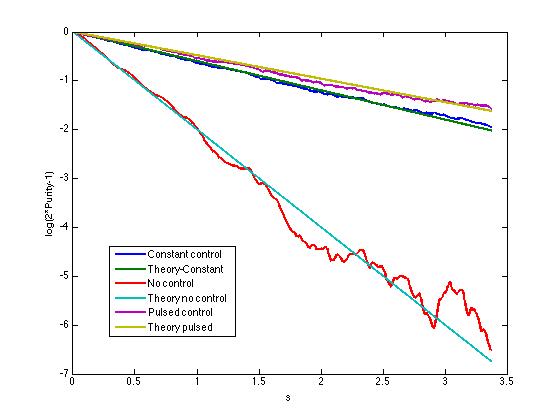}
\caption{ 
The logarithmic purity of the state,  $\log \,Tr(\rho(s)^2)$
as a function of the coarse grained time $s$,  Eq.~(\ref{s}) for stochastic evolutions and the corresponding Lindbladians. The weak coupling, Eq.~(\ref{e}),  is  $\e = 0.15$. The numerical grid $\Delta=1$ and  correlation time $\tau= 20\Delta$.  
The control parameter is $\omega_c \tau = 0.5\pi,$ and the stochastic averaging is done on an ensemble of $500$ runs.  }
\label{gammaT}
\end{center}
\end{figure}

A comparison between Lindbladian evolutions of sections~\ref{s:nc},\ref{s:bb},\ref{s:cc}
and  stochastic evolutions with Orenstein-Uhlenbeck process is shown in Fig.~\ref{gammaT}.   Three cases have been studied: no control, control by constant $H_0$ and Bang-Bang. 
The weak coupling parameter is $\e=0.15$ and the agreement is satisfactory. The numerical code is available upon request. 



\section{Summary}
We derived the Lindbladian  for controlled weakly stochastic evolutions  
both for small but finite $\e$ and in the limit $\e\to 0$ for stationary control.
Our results can be used to measure the power spectrum of the noise and
to test numerical algorithms for solving stochastic evolution.
Several examples  are studied in detail.

 \section*{Acknowledgment}
  We thank Nir Bar-Gil for drawing our attention to ref. \cite{lidar},   Amos Nevo and especially Ori Hirschberg and Martin Fraas for useful discussions. 
The research is supported by ISF, the EU Project DIADEMS, the Marie Curie Career Integration Grant (CIG) and the EU STReP project EQuaM.
 
 
\appendix



\section{Weak coupling expansion}\label{s:lin}

The purpose of  this appendix is to justify the estimate of Eq.~(\ref{U1}).
This requires a comparison of two different time orderings of the same exponent.
Let us first ignore the ordering and consider
\be 
e^x =\sum \frac {x^n} {n!}, \quad  x=-\int_0^t ds \, \C{G}(s)
\ee
The Taylor series for the exponent $e^x$ is dominated by terms of order $n=O(x)$.
In our case this gives $n\sim x\sim H^2 J\tau t\sim\e^2 t/\tau$.
Writing the n-th term in the expansion 
\begin{align}\label{xx}
\frac {x^n}{n!}= \int_{0\le s_1\le\dots s_n\le t} \prod_{i=1}^n \C{G}(s_i) \, ds_i 
\end{align}
we conclude that typically $s_{i+1}-s_i=O(t/n)=O(\tau/\e^2)\gg\tau$.

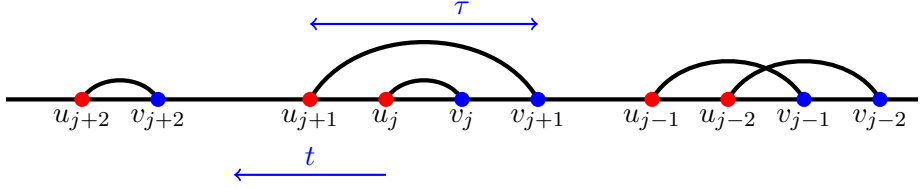
\begin{figure}[htbp]
\begin{center}
\begin{tikzpicture}
\draw [ultra thick] (0,0) -- (12,0);
\draw  [ultra thick] (1,0) to  [out=60, in=120]  (2,0);
\draw  [ultra thick] (5-1.,0) to  [out=60, in=120]  (8-1.,0);
\draw  [ultra thick] (6-1.,0) to  [out=60, in=120]  (7-1.,0);
\draw  [ultra thick] (5+3.5,0) to  [out=60, in=120]  (7+3.5,0);
\draw  [ultra thick] (6+3.5,0) to  [out=60, in=120]  (8+3.5,0);
\path [fill=red] (1,0) circle [radius=0.1];
\path [fill=blue] (2,0) circle [radius=0.1];
\path [fill=red] (5-1.,0) circle [radius=0.1];
\path [fill=red] (6-1.,0) circle [radius=0.1];
\path [fill=blue] (7-1.,0) circle [radius=0.1];
\path [fill=blue] (8-1.,0) circle [radius=0.1];
\path [fill=red] (5+3.5,0) circle [radius=0.1];
\path [fill=red] (6+3.5,0) circle [radius=0.1];
\path [fill=blue] (7+3.5,0) circle [radius=0.1];
\path [fill=blue] (8+3.5,0) circle [radius=0.1];
\node [below] at (1,0) {$u_{j+2}$};
\node [below] at (5-1.,0) {$u_{j+1}$};
\node [below] at (6-1.,0) {$u_{j}$};
\node [below] at (2,0) {$v_{j+2}$};
\node [below] at (7-1.,0) {$v_{j}$};
\node [below] at (8-1.,0) {$v_{j+1}$};
\node [below] at (5+3.5,0) {$u_{j-1}$};
\node [below] at (6+3.5,0) {$u_{j-2}$};
\node [below] at (7+3.5,0) {$v_{j-1}$};
\node [below] at (8+3.5,0) {$v_{j-2}$};
\draw [thick , blue] [<->] (5-1.,1) -- (8-1.,1);
\node [above, blue] at (6,1) {$\tau$};
\draw [thick , blue] [<-] (3,-1) -- (5,-1);
\node [above, blue] at (4,-1) {$t$};
\end{tikzpicture}
\caption{ $(u_j,v_j)$ make a dimer. The arc connecting the pair represents the short range interaction in Eq.~(\ref{yy}).  
In a typical configuration the distance between dimers is large $O(\tau/\e^2)$ and there is at most one dimer in 
an interval of size $\tau$. 
The dimer on the left is typical. The dimers  on the right   lead to the error.  The approximation
disregards the rare events.}\label{f:2}
\end{center}
\end{figure}

Next let us consider the possible time orderings.
The naive $\tilde{T}$ time ordering with respect to the argument of $\C{G}(s)$
implied by Eq.~(\ref{xx}) differs from the correct $T$-ordering because the relation between $\C{G}$ and $\C{H}$
\be\label{yy}
\C{G}(s_i)=-{1\over2}\int_{u_i>v_i} du_i dv_i j(u_i-s_i)j(v_i-s_i)\left(\C{H}(u_i)\C{H}(v_i)\right)
\ee
is non\blue{-}local in time: The ordering of $s_i$ does not guarantee the ordering of $(u,v)$.  The fact that $j(u-s)$ is fast decaying 
implies, however, that the nonlocality in time
is rather small $\sim\tau$.  When $s_{i+1}-s_i\gg\tau$  the   wrong ordering is almost the same as the correct one.

In order to estimate the error generated by using the $\tilde{T}$ ordering consider more closely the two orderings.
Each contribution to the exponent is given as in Eq.~(\ref{xx}) by some choice of ${0\le s_1\le\dots s_n\le t}$ and 
 we associate a choice of $u_i>v_i$ {to each $s_i$} as in Eq.~(\ref{yy}).
Typically $s_{i+1}-s_i\gg\tau\sim|u_i-s_i|,|v_{i+1}-s_{i+1}|$ and hence $v_{i+1}>u_i>v_i$ which    implies that the two ordering 
are equivalent.
If however there exists some $i$ for which $u_i>v_{i+1}$ then the two expressions do not coincide.

Consider  for example  the case where $u_{j+1}>u_{j}>v_{j+1}>v_{j}$ while all other points are at typical positions.
This will lead to an error term of the type
\begin{align}\label{dd}
\C{U}_{n,...j+2}\times
\int_{s_{j+1}>s_{j}}\hspace{-1cm}ds_{j}ds_{j+1}
\int_{u_{j+1}>u_{j}>v_{j+1}>v_{j}}\hspace{-2cm} du_{j}du_{j+1}dv_{j}dv_{j+1}
\C{H}(u_{j+1})\left[\C{H}(u_{j}),\C{H}(v_{j+1})\right]\C{H}(v_{j})\\
\times
j(s_{j}-u_{j})j(s_{j}-v_{j})j(s_{j+1}-u_{j+1})j(s_{j+1}-v_{j+1})
\times \C{U}_{j-1,...1}
\end{align}
Here $\C{U}_{n,...j+2}=\int_{u_{j+1}\le s_{j+2}\le\dots s_n\le t} \prod_{i=j+2}^n \C{G}(s_i) \, ds_i\;\;
      \C{U}_{j-1,...1}=\int_{0\le s_1\le\dots s_j\le v_j} \prod_{i=1}^{j-1} \C{G}(s_i) \, ds_i$
correspond to the (unitary) evolution before $t=v_j$ and after $t=u_{j+1}$.
The integrand in Eq(\ref{dd})is clearly fast decaying whenever its six integration variables are at inter-distance large 
compared to $\tau$.
It thus follows that the main contribution to the integral comes from a region of volume $\tau^5 t$.
The integral is thus at most\footnote{If $H_I(t)$ changes slowly in time\blue{,} then a tighter bound on the commutator is possible.} 
of order of $\tau^5 t j^4 \|{H}\|^4\sim \tau^3 t J^2  \|{H}\|^4=\e^4 t/\tau$.
Other nontypical cases (e.g. $u_{j}>u_{j+1}>v_{j+1}>v_{j}$) lead to error terms of a similar general form which  
again scale as $\e^4 t/\tau$.

The error terms we found are of the form $\int_0^t ds\, \C{U}(s,t)\Delta\C{G}(s)\C{U}(0,s)$ for some $\Delta\C{G}$ which is
quartic in $\C{H}$. This suggests defining an improved generator  as $\C{G}\mapsto\C{G}+\Delta\C{G}$.
We however    did not   pursue this direction here.

\section{ The spectrum of the super-operators of angular momenta}\label{a:spin}
The adjoint representation $ad(S)$ of a representation $S$ is constructed as the tensor product of $S$ with its 
dual (contragredient) representation $S^*$. Since $SU(2)$ has a single representaions in each dimension, it is
obvious that $S^*\simeq S$. It thus follows that
$$ad(S)=S\otimes S^*=S\otimes S=0\oplus 1\oplus 2\oplus......\oplus(2S)$$
The spectrum (including multiplicities) of various operators such as  $ad(S_z)$ and $\sum ad(S_j)ad(S_j)$ is then easily deduced
	\[
	Spect \big(ad(S_z)\big)= \bigcup_{j=0,\dots 2S} \{m| m=-j,\dots j\}
	\]

	\[
	Spect \left(\sum_{i=x,y,z}\big(ad(S_i)^2\big)\right)= \bigcup_{j=0,\dots 2S} \{j(j+1)| m=-j,\dots j\}
	\]

	\[
	Spect \left(\sum_{i=x,y}\big(ad(S_i)^2\big)\right)= \bigcup_{j=0,\dots 2S} \{j(j+1)-m^2| m=-j,\dots j\}
	\]
In particular the eigenvalue zero appears in $Spect\big(ad(S_x)^2+ad(S_y)^2\big)$ with trivial multilicity 1.
This last fact could also be deduced from Schur's lemma since by positivity $\big(ad(S_x)^2+ad(S_y)^2\big)\rho=0$
imply $ad(S_x)\rho=ad(S_y)\rho=0$ and hence also $ad(S_z)\rho=-i[ad(S_x),ad(S_r)]\rho=0$.


\section{Effective  control}
\label{effective}
In dynamical decoupling one is interested in making $\lin$ small at the price of strong control, $\omega_c\tau \gg 1$.   
Since $\tilde J(\omega)$ is small for large argument  and since the terms $\omega\neq 0$ in Eq.~(\ref{Fcg}) tend to be 
of order $\tilde{J}(\omega_c)$ the ``bad term" in $\lin$ is the one with $\omega=0$. We say that the control is ``effective" 
if $\tilde H_\alpha(\omega=0)=0$. The notion  is independent of $J_\alpha(u)$, which is often not known.


Consider first strong continuous controls.  
Let  $P_j(t)$ be the (instantaneous) spectral projections of $H_c$:
	\[
	H_c(t)=\omega_c\sum e_j(t) P_j(t)
	\] 
and suppose that $P_j(t)$ vary smoothly with $t$ and that the $e_j(t)$ do not cross. Then, by the adiabatic theorem, 
for $\omega_c$ large 
	\[
	H_\alpha^I(t)\approx \sum_{j,k} e^{i \omega_c\int_0^t(e_j(u)-e_k(u))du} P_j(t) H_\alpha P_k(t)
\underset {\omega_c\to\infty}{\longrightarrow}\sum_{j} P_j(t) H_\alpha P_j(t)
	\]
(in the sense of distributions.) It follows that the control is effective if, for all $t$,
\be\label{effective}
\sum_{j}P_j(t)H_\alpha P_j(t)=0
\ee
Bang-Bang at times $t_j$  is effective if  $\tilde H_\alpha(\omega=0)=0$, which is the case if $H_I(t)$ has zero average, i.e.
	\[
\forall\alpha,\;\;
	\sum_j (t_{j+1}-t_j) V(t_j) H_\alpha V(t_j)=0
	\]


\begin{thebibliography}{10}

\bibitem{alicki}
Robert Alicki and Mark Fannes.
\newblock Quantum dynamical systems.
\newblock {\em status: published}, 2001.

\bibitem{alicki-lidar}
Robert Alicki, Daniel~A Lidar, and Paolo Zanardi.
\newblock Internal consistency of fault-tolerant quantum error correction in
  light of rigorous derivations of the quantum markovian limit.
\newblock {\em Physical Review A}, 73(5):052311, 2006.

\bibitem{Ido}
Ido Almog, Yoav Sagi, Goren Gordon, Guy Bensky, Gershon Kurizki, and Nir
  Davidson.
\newblock Direct measurement of the system-environment coupling as a tool for
  understanding decoherence and dynamical decoupling.
\newblock {\em J. Phys. B}, 44:154006, 2011.

\bibitem{BreuerPet}
Heinz-Peter Breuer and Francesco Petruccione.
\newblock {\em The Theory of Open Quantum Systems}.
\newblock Oxford University Press, USA, 2007.

\bibitem{davies-markovian}
E~Brian Davies.
\newblock Markovian master equations.
\newblock {\em Communications in mathematical Physics}, 39(2):91--110, 1974.

\bibitem{davies}
Edward~Brian Davies.
\newblock Quantum theory of open systems.
\newblock 1976.

\bibitem{morigi}
Berthold-Georg Englert and Giovanna Morigi.
\newblock Five lectures on dissipative master equations.
\newblock In Andreas Buchleitner and Klaus Hornberger, editors, {\em Coherent
  Evolution in Noisy Environments}, volume 611 of {\em Lecture Notes in
  Physics}, pages 55--106. Springer Berlin Heidelberg, 2002.

\bibitem{fraas}
M.~{Fraas}.
\newblock {Adiabatic theorem for a class of quantum stochastic equations}.
\newblock {\em ArXiv e-prints}, July 2014.

\bibitem{lidar}
Goren Gordon, Gershon Kurizki, and Daniel~A Lidar.
\newblock Optimal dynamical decoherence control of a qubit.
\newblock {\em arXiv preprint arXiv:0804.2691}, 2008.

\bibitem{GoriniKossakowski}
V.~Gorini, A.~Kossakowski, and E.C.G. Sudarshan.
\newblock Completely positive dynamical semigroups of {$N$}-level systems.
\newblock {\em J. Math. Phys.}, 17(5):821--825, 1976.

\bibitem{James_effe}
Daniel F.~V. James and Jonathan Jerkel.
\newblock Effective hamiltonian theory and its applications in quantum
  information.
\newblock {\em Canadian Journal of Physics}, 85:625--632, 2007.

\bibitem{kurizki}
A.~G. Kofman and G.~Kurizki.
\newblock Unified theory of dynamically suppressed qubit decoherence in thermal
  baths.
\newblock {\em Phys. Rev. Lett.}, 93:130406, Sep 2004.

\bibitem{Lindblad}
G.~Lindblad.
\newblock On the generators of quantum dynamical semigroups.
\newblock {\em Comm. Math. Phys.}, 48:119--130, 1976.

\bibitem{paola1}
Carlos Meriles, Liang Jiang, Garry Goldstein, Jonathan Hodges, Jeronimo Maze,
  Mikhail Lukin, and Paola Cappellaro.
\newblock Imaging mesoscopic nuclear spin noise with a diamond magnetometer.
\newblock {\em THE JOURNAL OF CHEMICAL PHYSICS}, 133:124105, 2010.

\bibitem{bargil1}
Y~Romach, C~M\"uller, T~Unden, L.~J. Rogers, T~Isoda, K.M. Itoh, M.~A. Markham,
  Stacey, J~Meijer, S~Pezzagna, B~Naydenov, L.P. McGuinness, N.~Bar-Gill, and
  F~Jelezko.
\newblock Nuclear magnetic resonance spectroscopy on a (5-nanometer)3 sample
  volume.
\newblock {\em Arxiv}, 1404:3879, 2014.

\bibitem{salgado}
D.~{Salgado} and J.~L. {Sanchez-Gomez}.
\newblock {Lindbladian Evolution with Selfadjoint Lindblad Operators as
  Averaged Random Unitary Evolution}.
\newblock {\em eprint arXiv:quant-ph/0208175}, August 2002.

\bibitem{Milburn1}
S.~Schneider and G.~J. Milburn.
\newblock Decoherence and fidelity in ion traps with fluctuating trap
  parameters.
\newblock {\em Phys. Rev. A.}, 59:3766, 1999.

\bibitem{LebowitzSpohn}
Herbert Spohn and Joel~L. Lebowitz.
\newblock Irreversible thermodynamics for quantum systems weakly coupled to
  thermal reservoirs.
\newblock {\em Advances in Chemical Physics}, pages 109--142, 1978.

\bibitem{joerg1}
T~Staudacher, F~Shi, S~Pezzagna, J~Meijer, J~Du, C~Meriles, F~Reinhard, and
  J~Wrachtrup.
\newblock Nuclear magnetic resonance spectroscopy on a (5-nanometer)3 sample
  volume.
\newblock {\em Science}, 399(6119):561, 2013.

\bibitem{Szczygielski}
K.~{Szczygielski}.
\newblock {On the application of Floquet theorem in development of
  time-dependent Lindbladians}.
\newblock {\em ArXiv e-prints}, March 2014.

\end{thebibliography}
\end{document}